# 1. Introduction

Since their first introduction by Schoenbach et al., in 1995, micro-plasmas have gained the interest of the plasma scientific community. In 1997, the first micro-cavities were manufactured using microtechnology techniques [1]. Spatially confining the plasma to sub-millimetric dimensions is an interesting approach to the generation and the maintenance of stable glow discharges at atmospheric pressure. These stable micro-plasmas can be generated in various geometries [2]. The most studied are: micro-jets, inverted pyramidal structures, cathode boundary layer discharges (CBL) and micro hollow cathode discharges (MHCD). A relevant parameter characterizing these microplasmas is their high surface-to-volume ratio, thus indicating the importance of wall effects on the plasma. According to Kushner [3], they provide a higher degree of regulation of gas temperature by thermal conduction.

We have recently reported the fact that a microdischarge can be forced to operate in an abnormal glow regime by reducing the cathode surface area [4]. This can allow one to initiate the plasma in several micro-reactors simultaneously with a single ballast resistance. In this paper, we present simulation results and experimental measurements that help one to understand how limiting the cathode surface area affects the discharge properties. Several parameters were varied or measured. First, I – V characteristics were both measured and simulated as the cathode area was increasingly limited. Gas temperature, sheath thickness and electron density were also studied experimentally and by simulation. Finally a hysteresis effect, also reported in [4], was investigated further.

# 2. Experimental setup and description of the model

## 2.1. Experimental setup

The MHC device used for the experiment is a Ni:Al$_2$O$_3$:Ni sandwich structure having one hole of ~260 µm diameter and a 250 µm thick dielectric. The electrodes consist of approximately 8 µm thick nickel layers deposited using an electro-plating technique. The hole was made by laser drilling. The dielectric area is 1×1 cm$^2$ and the electrode area is 0.9 × 0.9 cm$^2$.

A dielectric layer (Kapton tape) was used to partially cover the cathode surface and thereby control the cathode surface area. The Kapton layer is effective at reducing the cathode area because we use a DC power supply. This Kapton layer is modelled using two characteristic dimensions: the diameter of the hole in it (D$_{LIM}$) and its thickness (t$_{LIM}$) as presented in Figure 1. For the simulation, three values of D$_{LIM}$ (500 µm, 1 mm and no limiting layer) and two values of t$_{LIM}$ (60 µm and 1 mm) were used. This resulted in four characteristic geometries, shown in Figure 1, which will be discussed in the following sections. The first geometry is a typical MHC device without any limiting layer. The second is a MHC device covered by a dielectric layer having D$_{LIM}$ = 1 mm and t$_{LIM}$ = 60 µm; the third has D$_{LIM}$ = 500 µm and t$_{LIM}$ = 60 µm and the final has D$_{LIM}$ = 1 mm and t$_{LIM}$ = 1 mm.

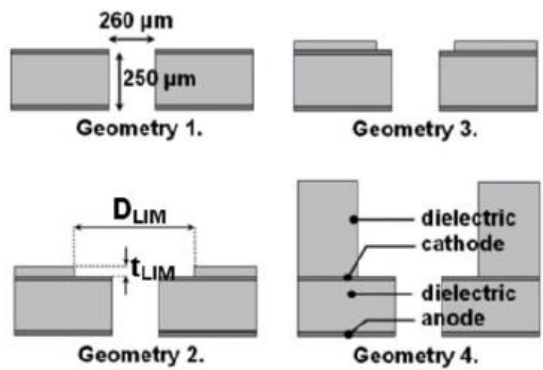

*Fig. 1. Schematics of the modelled environments showing MHC devices both without a dielectric (limiting) layer covering the cathode (Geometry 1), and with 3 different limiting layers: D$_{LIM}$ = 1 mm and t$_{LIM}$ = 60 µm (Geometry 2), D$_{LIM}$ = 500 µm and t$_{LIM}$ = 60 µm (Geometry 3) and D$_{LIM}$ = 1 mm and t$_{LIM}$ = 1 mm (Geometry 4).*

A 0–2500 V DC (300 W) power supply was used to generate the micro-discharge. A 39 kΩ ballast resistor in series with the micro-discharge regulates the input current and another 1 kΩ resistor connected between the anode and ground permits an easy measurement of the current using an oscilloscope. The MHC device can be approximated using an equivalent electric circuit as detailed in a paper by Aubert et al. [5].

The micro-discharge can operate in the pressure range of 100 to 1000 Torr, as measured by an absolute pressure transducer (MKS 626A Baratron gauge). Operating in helium allows one to ignite the discharge at voltages lower than those needed to ignite in an air ambient. The chamber is evacuated using a primary pump and refilled with helium 3 times in order to minimize contaminants in the gas; however there are always some trace levels of nitrogen (<0.5%). This background has no appreciable impact on the measured V − I curves. According to our experiments, independently the helium's quality injected in the chamber, the value of the breakdown voltage was always the same and the normal glow voltage was unchanged. Most of the ionization is produced in the high field sheaths by direct processes. Penning ionization in the low field plasma regions could also be a factor contributing to the ionization balance when a significant concentration of nitrogen is present. However, based on the fact that little difference is observed in the experimental V −I characteristic with increasing nitrogen concentration up to 1%, it appears that the Penning contribution to the ionization balance is small. The background of 5000 ppm of nitrogen was therefore a very convenient way for performing measurements of gas temperatures. By optical emission spectroscopy, we analysed the rovibrationnal structure of the 1–3 and 0–2 bands of the $N_2$ second positive system [6,7]. The nitrogen emission spectra were recorded with a 1 m focal length monochromator (HR1000) having a 2400 lines/mm grating and a CCD camera detector. To evaluate the electron density, we have recorded the spectral profile of the $H_\beta$ line emission from the microdischarge [8–10]. This emission is also due to trace impurities in the micro-discharge chamber. The determination of the electron density from the Hβ line width requires a very accurate spectral resolution. Consequently we used a 2 m focal length monochromator (SOPRA) having a 1200 groves/mm grating working in the 3$^{rd}$ diffraction order and the same CCD camera. This monochromator provides a spectral dispersion of 0.97 pm/pixel on the CCD camera. With the 100 µm entrance slit, the spectral resolution was 7 pm, much smaller than the widths of the $H_\beta$ line profiles.

## 2.2 Description of the model

The model software used to simulate our micro-discharges was developed by Boeuf and Pitchford at the Laplace Laboratory (Toulouse, France) [11]. This model allows the simulation of glow discharges in 2 dimensions.

### 2.2.1. Computational domain and fundamental equations

The computational domain, shown in Figure 2, is cylindrically symmetric with dimensions of 4 mm height by 2.5 mm radius. The characteristic dimensions of the simulated MHC device are nominally the same as in our experiments (electrode thickness of 8 µm, dielectric thickness of 250 µm, hole diameter of 260 µm). We used a non-uniform grid of 90 by 120 nodes, to provide computational resolution in the cathode sheath regions and inside the cavity. This grid is assigned to a cylindrical coordinate system and the computational volume is assumed to be bounded by a dielectric layer at a distance sufficiently far so as to not influence the results.

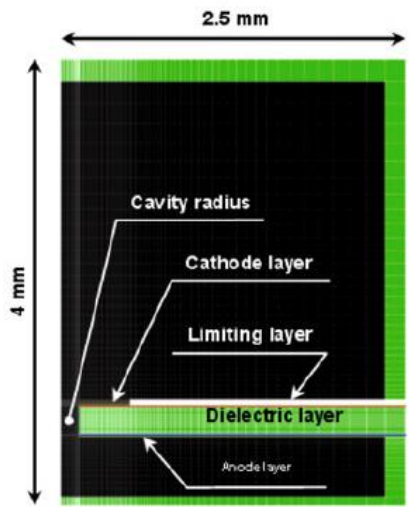

*Fig. 2. Computational domain of the MHC device. The axis of cylindrical symmetry is on the left and the computational grid is superimposed.*

The model solves the fluid equations in the drift-diffusion approximation. These equations ensure the continuity of the electron and ion transport and are coupled equation and a heat conduction equation complete the fluid model [11]. The model results are obtained by assuming that the discharge is powered by a voltage source through a series resistor. The value of the series resistor and the source voltage are adjusted to attain the desired current. Transport coefficients and rate coefficients for helium were determined by using the freeware solver BOLSIG+, which solves the Boltzmann equation from collision cross-section data [12]. The cross sections used in the Boltzmann calculation were taken from the compilation of AV Phelps [13].

**2.2.2. Assumptions**

The model was simplified by supposing that all the ionization is due to electron impact ionization of atoms in the ground state and thereby neglecting all reactions involving metastables. This simplification is somewhat justified because we are focusing on the sheath regions where the metastable densities should be negligible and where most ions should be atomic ions. Therefore, only ground state helium atoms, electrons and one species of positive ion were considered. The gas heating source term includes both ion and electron currents, but the ion contribution is by far dominant. Computing the gas heating caused by ion current in the sheath would require a Monte Carlo simulation of the ions and of the fast neutral species in the sheath. In our simple model, the fraction of the ion energy responsible for the gas heating was set to 25%. This estimation of 25% is given by Revel et al. [14], who showed that, in the case of a glow discharge in argon over a range of conditions, 75% of the total ion energy in the cathode sheath is deposited directly at the cathode and 25% is converted to gas heating in the sheath. In addition, the cathode and dielectric surfaces were assumed to be at a constant temperature of 300 K. A thermal boundary layer is included as a boundary condition in the calculation [15].

# 3. Results

## 3.1. Breakdown mechanism

In a geometry where the electrodes are in the form of two parallel planes, the gas breakdown occurs along the interelectrode distance [16], whereas in the case of a classical hollow cathode, this phenomenon depends on the radius of the cavity, since pendulum electrons are involved in the charge generation processes [17]. For our geometry (microhollow cavity), one could wonder which characteristic dimension (diameter or length of the cavity) dominates. To elucidate the role of each dimension in the breakdown process, we tested a micro-device having a single cavity, as presented in the inset of Figure 3. Its length is as much as 400 µm and its circular openings present two different diameters: 90 µm on the lower surface and 190 µm on the upper surface. By assuming the breakdown to only occur on the inter-electrode distance, we plotted two Paschen curves in Figure 3: the first curve was obtained with the lower surface acting as cathode (black squares) and the second curve with the upper surface acting as the cathode (red open circles). The two Paschen curves align to within experimental precision over the range of pressures between 10 and 400 Torr without any dependence on which side was used as cathode.

Therefore, the Paschen curves do not depend upon the cavity diameter at the cathode as should be the case if a hollow cathode effect were dominant. Even though the geometry is that of a micro-hollow cathode, its diameter does not seem to have any impact on the breakdown mechanisms. The electrons emitted from the cathode do not execute a pendulum motion; but instead are accelerated towards the anode and thus the anode-cathode distance is the relevant parameter for the breakdown. The cathode is too thin at 8 µm for pendulum electrons to be important. The breakdown phenomenon likewise does not depend on the spatial limitation of the cathode surface. However, the steady state regime does, as we will discuss now.

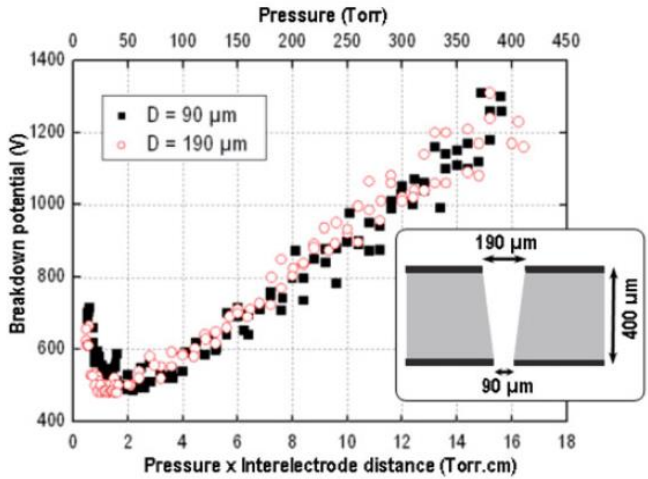

*Fig. 3. Paschen curves of a micro-hollow cathode and characteristic dimensions of the micro-cavity tested.*

## 3.2. Steady state regime of the micro-discharge

### 3.2.1. Steady state regime for an unlimited cathode area

The experimental V – I curve of a micro-discharge operating at 100 Torr in helium is shown in Figure 4 (black squares) for an unlimited cathode area. The characteristic dimensions of the micro-cavity are those of geometry 1 in Figure 1. If the current was limited by the power supply to lower than 1 mA the micro-discharge current would switch on and off at a nearly steady frequency. This is the self-pulsing regime studied by Aubert et al. [5] and is not addressed further here. The line between data points A and B is drawn in order to help guide the eye. It shows the jump in current flowing through the MHCD when the glow ignites. The slope of this line is negative and set by the fact that the voltage decrease across the MHC device is now dropped across the external electrical circuit (the ballast resistance of 40 kΩ). Once the discharge has ignited, it operates in a normal glow regime. The discharge voltage remains essentially constant with increasing current because the cathode surface area covered by the discharge can increase instead. This is seen over the full range of the V –I curve in Figure 4, from 4 to 28 mA. The maximum current of 28 mA in Figure 4 is not a limitation due to the micro-device geometry. Currents as large as 40 mA, could flow through the MHCD without breaking the device by thermal stresses. Such current levels correspond to a power of ~7 W for a single hole.

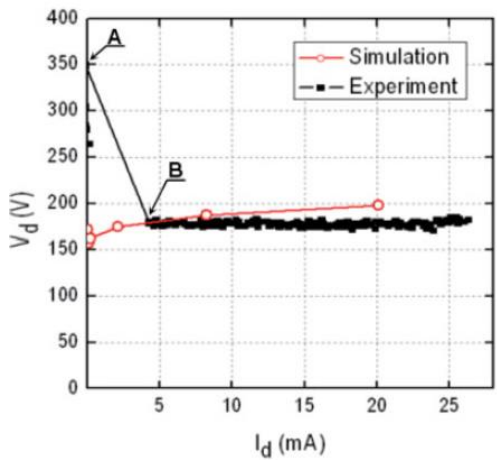

*Fig. 4. Experimental (black squares) and simulated (red circles) V – I curves for a micro-discharge generated in helium at 100 Torr. The microcavity has a diameter of 260 µm and the cathode surface area is not limited. The solid line between points A and B helps to guide the eye showing the current increase and discharge voltage decrease which occurs at ignition. The simulation results were obtained for a value of the secondary electron emission coefficient of 0.3.*

The simulation of the micro-discharge at the same experimental conditions gives the V – I curve marked by open circles in Figure 4. Each data-point represents a stationary solution for the micro-discharge model. Good agreement between simulation and experiment is obtained for a secondary electron emission coefficient ($\gamma_{se}$) value of 0.3, a value consistent with secondary electron emission induced by bombardment of atomic helium ions on metal surfaces [18]. As mentioned above, the simulated V – I characteristic is obtained by varying the source voltage and the ballast resistance. The value of the ballast resistance needed to simulate the low current data points was quite large (R > 2 MΩ). In addition, there are no experimental data points at low currents (between points A and B). As a result, we cannot compare the simulation with the experiment in that range of small currents. For currents larger than ∼4 mA, the simulated and experimental results should be directly comparable. The experimental V –I curves have a flat slope whereas the simulated V –I curve has a very slight positive slope. The reason is that, experimentally, the surface of the cathode was not limited. However, in the simulations, it is necessary to define a computational domain, thus limiting the volume in which the micro-discharge was operating and causing a slight positive slope in the V – I curve for high current (confirmed by changing the cathode diameter in the simulations).

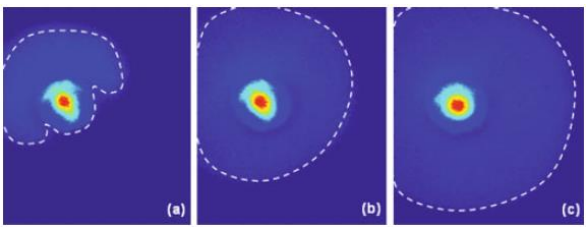

*Fig. 5. ICCD Pictures of the micro-discharge optical emission intensity on the cathode side (relative scale). The plasma increasingly spreads over the cathode surface as discharge current increases. (a) $I_d$ = 5 mA; (b) $I_d$ = 9 mA; (c) $I_d$ = 20 mA. The diameter of the MHC circular cavity is 260 µm.*

During the normal glow regime, the micro-discharge expands over the cathode surface, as shown in the ICCD pictures in Figure 5. In this regime, the discharge voltage remains approximately constant (∼175 V), and the increase in Id is due to the spread of the micro-discharge over the cathode surface where the discharge appears to be uniform. The area of the spread is almost linearly dependent on the current level and corresponds to an average of 30 mA/mm2.

We have determined the gas temperature by fitting the experimental profiles of the 1–3 band of the nitrogen second positive system. The gas temperature was the adjusting parameter of the spectral simulation code. The vibrational and rotational constants of the upper [19] and lower [20] levels of the transition and the transition probabilities of rotational lines of different branches [21] were taken from the literature and had fixed values in the fitting code, as was the experimental parameter specific of the apparatus function of the spectrometer. As an example, Figure 6a shows a fit of the experimentally recorded 1–3 band of $N_2$ with a simulated spectrum. The estimated uncertainty on the gas temperature is less than 20 K.

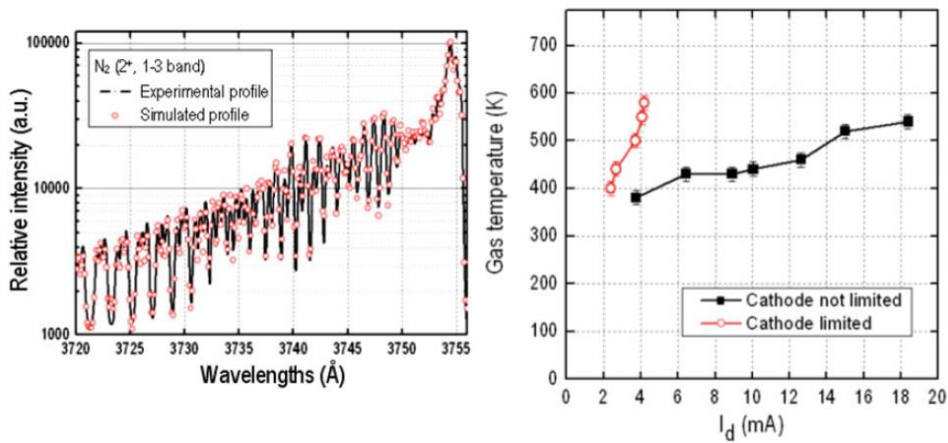

*Fig. 6. (a) Fit on the experimental spectra of the 1–3 band of the second positive system of nitrogen at 100 Torr and 15 mA providing $T_g$ = 520 K. (b) Gas temperature versus current at 100 Torr in helium for $D_{LIM}$ = 1 mm and for non limited cathode surface.*

The experimentally determined gas temperature in the micro-discharge is plotted versus current for two cases in Figure 6b: (i) without limitation of the cathode and (ii) for $D_{LIM}$ = 1 mm. The gas temperature in the microdischarge can reach 540 K at $I_d$ = 18 mA when the cathode area is not limited. When the cathode surface is limited, the largest attainable discharge current is reduced to only 4 mA but the gas temperature reaches 600 K. These two curves show that by reducing the cathode area, the micro-discharge can reach higher gas temperatures at lower currents. This is expected because the current density in the cathode sheath region is forced to become larger.

In the steady state regime, the electron number density can be measured inside the cavity by determining the Stark broadening of the Hβ line. Several line profiles obtained for currents ranging from 2 mA to 25 mA, are presented in Figure 7a for the micro-discharge operating in helium at 100 Torr. Increasing the current leads to: (i) a stronger Stark effect which results in the broadening of the profile; (ii) a higher gas heating by Joule effect and thus a higher gas temperature, which according to the relation $p = nk_BT_g$ = 100 Torr implies a lower neutral density and thus a lower Van der Waals broadening; and (iii) a larger Doppler width with increasing temperature.

The Hβ line profile is a Voigt profile, which mathematically is the convolution of a Gaussian function by a Lorentzian function. The Stark broadening – represented by the Lorentzian function – influences only the wings of the profile. As a consequence, this broadening can be estimated by fitting a Lorentzian curve to the Lorentzian profile wings alone. Then, the electron number density can be calculated according to the method described in [9]. This is what has been plotted versus Id in Figure 7b for 100 Torr, 400 Torr and 750 Torr gas pressures. At 100 Torr, ne increases from $1.4 \times 10^{14}$ cm$^{-3}$ (at 2 mA) to $3.6 \times 10^{14}$ cm$^{-3}$ (at 25 mA). The electron densities deduced from Stark broadening are also pressure dependent. The values of ne at atmospheric pressure are found to be larger than those at 100 Torr for all discharge currents.

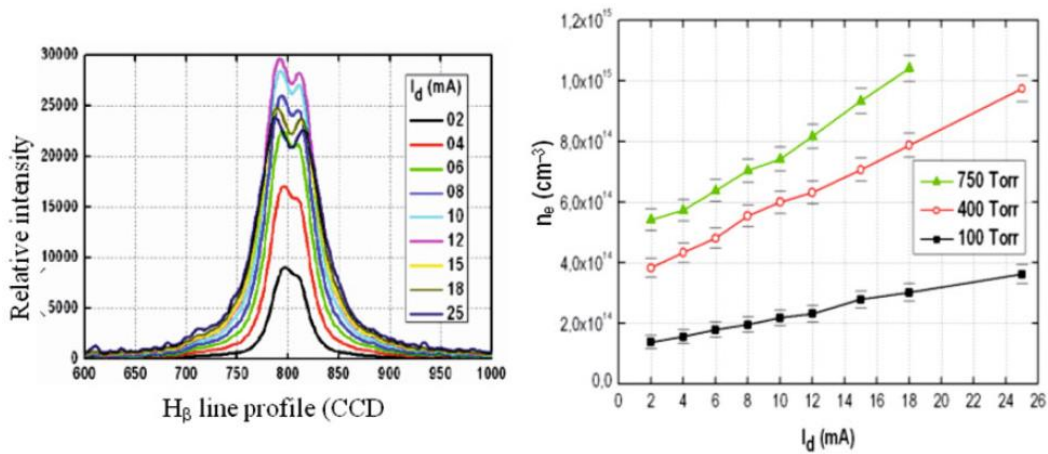

*Fig. 7. (a) Hβ line profile in helium at 100 Torr for $I_d$ = 2–25 mA. (b) Electron number densities versus the operating current for a single micro-discharge working at 100, 400 and 750 Torr in helium. The cavity diameter is approximately 260 μm.*

**3.2.2. Steady state regime for a limited cathode area**

The cathode surface area was reduced by covering it with a limiting layer whose characteristic dimensions are its thickness ($t_{LIM}$ = 60 µm) and its opening diameter ($D_{LIM}$ = 1 mm). The micro-discharge is supplied by a linearly increasing ramp of direct current, and is operated in helium at 100 Torr. Its experimental V – I curve is plotted in Figure 8 (black squares). Once ignited, the micro-discharge current increased from 3 to 9.5 mA and operates in an abnormal glow regime characterized by a positive slope of the V – I curve. The voltage reaches a maximum value of 380 V at the maximum test current of 9.5 mA. The simulated V –I curve (open circles) matches the experimental V – I curve (black dots) for $\gamma_{se}$ = 0.2. Simulation results for $\gamma_{se}$ = 0.25 and $\gamma_{se}$ = 0.3 are also shown in Figure 8 for comparison with results shown in Figure 4.

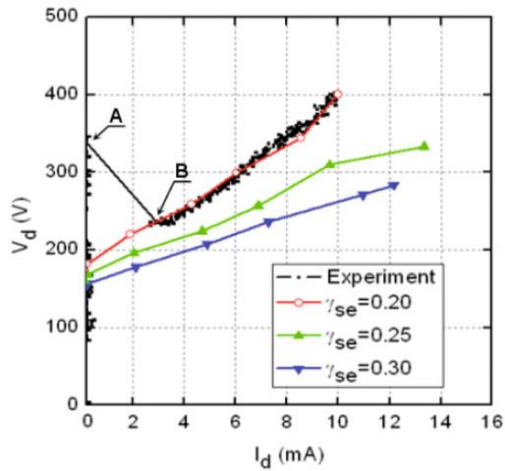

*Fig. 8. Experimental and simulated V –I curves for $D_{LIM}$ = 1 mm. The micro-discharge is operating in helium at 100 Torr and the best fit is obtained for $\gamma_{se}$ = 0.2. The slope of the load line between points A and B corresponds to the value of the experimental ballast resistor: 40 kΩ.*

## 3.3. Influence of the opening area on the micro-discharge properties

The results obtained for various cathode surface areas (opening diameters) are presented in this section. In all cases, the cathode surface limiting layers made of Kapton having a thickness of 60 µm.

**3.3.1. Influence of cathode surface area on the V – I curve**

The influence of the opening diameter ($D_{LIM}$) can be studied by plotting V – I curves, with different DLIM on the same graph. We used a micro-device whose cavity diameter is 250 µm and inter-electrode distance is 250 µm. This micro-device was operated in helium at 750 Torr and supplied by a triangle wave DC voltage ramp with a period of 40 s. Three different limiting layers were successively tested on cathode of this micro-device. Their opening diameters were $D_{LIM}$ = 2.5 mm, $D_{LIM}$ = 1.5 mm and $D_{LIM}$ = 0.6 mm. The corresponding V –I curves are plotted in Figure 9. The small arrows on the $D_{LIM}$ = 0.6 mm curve indicate the direction of travel around the hysteresis curve as the voltage first increases in time (arrows 1, 2, 3) and then decreases (arrows 4 and 5). The experimental V –I curves show hysteresis that might be attributable to a thermal relaxation time or other slow process. A thermal relaxation time will depend on the thermal conductivity of the gas and of the micro-device material.

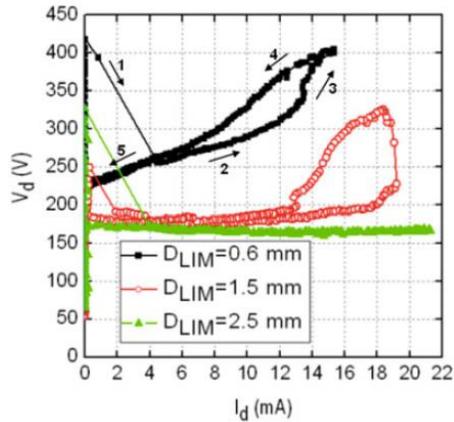

Fig. 9. Influence of $D_{LIM}$ on the slope of the abnormal glow regime, for a micro-discharge operating in helium at atmospheric pressure.

For $D_{LIM}$ = 2.5 mm, the voltage remains nearly constant from 0 to 21.5 mA: no abnormal glow regime is observed because the opening area is large and does not limit the plasma spread. In contrast, the slope clearly increases with current (especially from 13 to 19 mA) for $D_{LIM}$ = 1.5 mm and has an even larger positive slope for $D_{LIM}$ = 0.6 mm ranging over the full V – I curve from 0 to 15 mA. These three curves clearly show that a reduction of the cathode surface area causes an increase in the discharge voltage, allowing the micro-discharge to enter into an abnormal glow regime.

To better understand this transition from normal to abnormal glow regime, we focused our attention on the results from the simulation of the potential distribution in the cathode sheath regions. Simulations were carried out for the same conditions as in the experiments: helium at 100 Torr for a current of 5 mA, in the three following configurations:
- No limiting layer
- $D_{LIM}$ = 1 mm ($t_{LIM}$ = 60 µm);
- $D_{LIM}$ = 500 µm ($t_{LIM}$ = 60 µm).

The potential profiles for these three configurations, presented in Figure 10 for a subset of the computational volume, show that a reduction of $D_{LIM}$ causes an increase in the the discharge voltage and that most of the voltage drop is in the cathode sheath region. For instance, the cathode sheath voltage located in the plane at middle of the cathode layer's thickness are: 160 V without limitation (corresponds to Fig. 4 where the total discharge voltage at 5 mA is ∼180 V), 185 V for $D_{LIM}$ = 1 mm and 330 V for $D_{LIM}$ = 500 µm. The simulated sheath voltage is therefore two times larger for $D_{LIM}$ = 500 µm than for the unlimited case.

These three simulation results also indicate an increasing voltage along the central axis (Δ) of the cavity due to the current flow that is relatively independent of the value of $D_{LIM}$. The axial voltage profiles are shown in Figure 11, in which we distinctly observe three linear curves with the same slope and thus a same voltage drop (almost 20 V). We can also estimate at almost 800 V/cm the on-axis electric field strength, corresponding to a reduced electric field strength, E/p, of 8 V/cm/Torr. This region can be likened to the positive column of the micro-discharge, although the neglect of metastables in the model leads to an overestimate of the on-axis field because we do not include step-wise ionization. We can nevertheless conclude that the spatial limitation of the cathode surface has little impact on the axial electric field inside the cavity; its only influence is on the cathode sheath voltages. This is consistent with previous model predictions of the V –I characteristic of a MHCD for varying dielectric thickness [22].

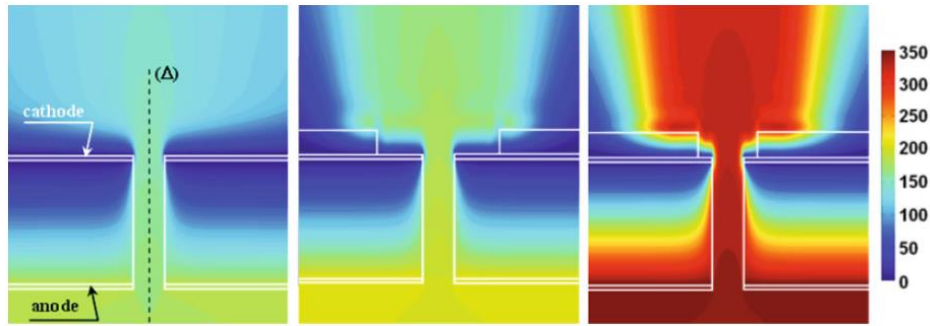

Fig. 10. Calculated equipotential profiles (with z and r axes having different scales) showing the micro-plasma region at 5 mA for limiting layers having a thickness of 60 µm. The glows were simulated to be in helium at 100 Torr. (a) No cathode area limitation; (b) $D_{LIM}$ = 1 mm; (c) $D_{LIM}$ = 500 µm.

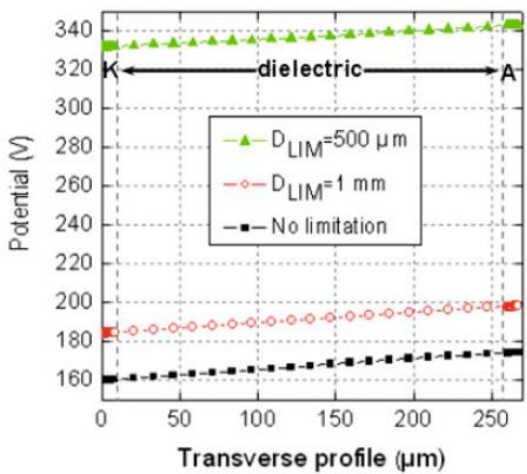

Fig. 11. Simulation results for the potential profile along the central axis of the cavity for a cathode surface: (a) not limited, (b) $D_{LIM}$ = 1 mm and (c) $D_{LIM}$ = 1 mm.

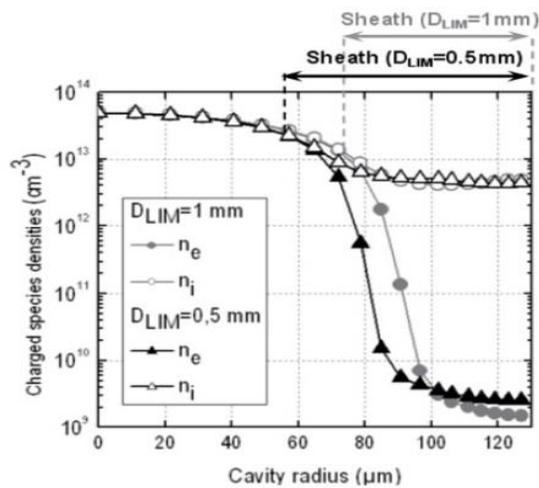

Fig. 12. Simulated profiles of charged species densities along the radius of the cavity. The cathode surface is at 130 µm and the discharge axis is at zero. Two cases are considered at 100 Torr in helium: $D_{LIM}$ = 1 mm and $D_{LIM}$ = 500 µm, both for $I_d$ = 5 mA.

**3.3.2. Influence on the charged species densities**

Figure 12 shows the radial profiles of the electron and ion densities ($n_e$ and $n_i$) at an axial position corresponding to the middle of the cathode's thickness for two limiting cases:

- (i) In the case $D_{LIM}$ = 1 mm, $n_e$ and $n_i$ gradually decrease with increasing radial coordinate from the hole axis at 0 µm to the cathode's edge (surface) at 130 µm. The electron and ion densities are nearly equal (quasineutrality) in the bulk plasma volume between 0 and ~70 µm where the plasma density is about $4 \times 10^{13}$ cm$^{-3}$. The sheath region has a thickness of about 60 µm in which the electron density drops rapidly and where the ion density is roughly constant at about $5 \times 10^{12}$ cm$^{-3}$.
- (ii) In the case $D_{LIM}$ = 0.5 mm, Figure 12 shows that the bulk plasma volume is now localized inside a slightly smaller region because the sheath thickness is somewhat larger: its value is now about 65 µm. The ion density becomes approximately constant beyond 90 µm (~ $5 \times 10^{12}$ cm$^{-3}$). The constant (positive) space charge in the sheath region is consistent with a linearly decreasing electric field strength with increasing distance from the cathode surface.

To sum up, both the simulation and the experiments show the operating voltage increases when the cathode surface area is decreased for a constant discharge current of 5 mA. The simulations show that the cathode sheath length inside the cavity changes only slightly with an increase in the total discharge voltage from 180 to 330 V (corresponding to a decrease of the diameter of the exposed electrode surface from 1 mm to 500 µm). Because the thickness of the cathode film is very small compared to the sheath dimensions of the sheath in Figure 12, it is not possible to treat the sheath as if it were one-dimensional. As a result, it is not possible to make use of analyses based on one-dimensional sheath structures to help understand the results of the simulation either.

**3.3.3. Influence on gas temperature**

The calculated gas temperatures are in reasonable agreement with the experimental results shown in Figure 6 for similar conditions. For an unlimited cathode area (Fig. 13a) the gas temperature reaches a maximum of about 380 K. We find a modest increase in the peak gas temperature to about 430 K when the micro-device is covered by a limiting layer with $D_{LIM}$ = 1 mm (Fig. 13b). The temperature increase is significantly higher for for $D_{LIM}$ = 0.5 mm (Fig. 13c) with the peak value at about 650 K.

Gas heating is mainly due to the ion current in the sheaths and the gas temperature increases with increasing local current density. There is a decrease of about a factor of four in the cathode surface area between Figures 13b and 13c, and since the current is constant, the current density and thus the gas temperature higher for the geometry of Figure 13c.

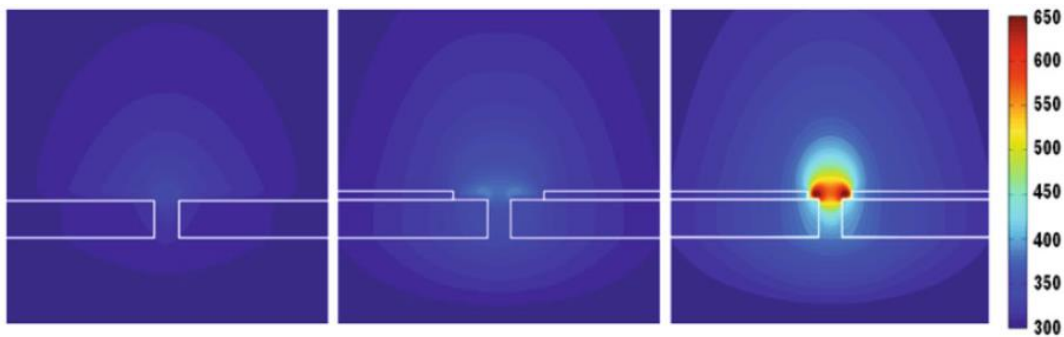

*Fig. 13. A zoom of the gas temperature (Kelvin) in the cathode region at 5 mA: (a) no spatial limitation of the cathode; (b) $D_{LIM}$ = 1 mm; (c) $D_{LIM}$ = 500 µm. In the three cases, $t_{LIM}$ = 60 µm and the micro-discharge is working in helium at 100 Torr.*

## 3.4. Influence of the limiting layer thickness on the micro-discharge properties

We also simulated the influence of the limiting layer thickness on the electrical and thermal properties of the microdischarge. To simplify our approach, we only considered two thicknesses: 60 µm (Fig. 1, Geometry 2) and 1 mm (Fig. 1, Geometry 4) with the same opening diameter of 1 mm. The V – I curves relative to these two configurations are plotted in Figure 14, in which we also report the case of an unlimited cathode area operated at the same conditions. The discharge voltage increases linearly from 160 V (at 0.01 mA) to 285 V (at 12 mA) for $t_{LIM}$ = 60 µm (open circles). The same trend is observed for the curve at $t_{LIM}$ = 1 mm (green triangles) along with a constant offset of ∼20 V. This offset value is very small by comparison to the operating voltages (>160 V). As a consequence, the electrical properties remain almost unchanged even if the value of $t_{LIM}$ is greatly changed (by a factor of 16 in this case). This indicates that it is the limiting of the cathode area by an insulator that causes these effects rather than the thickness of the insulator itself.

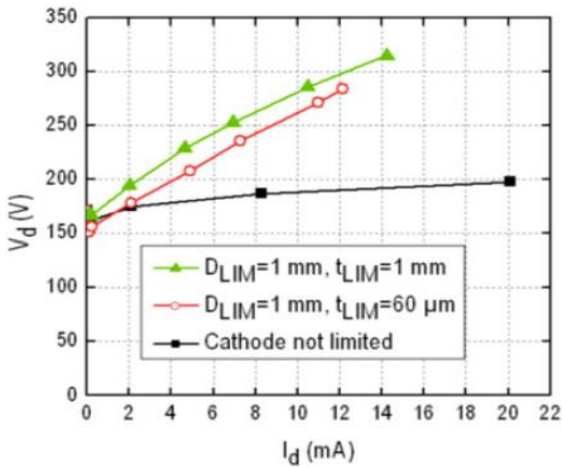

*Fig. 14. Simulated V – I curves of a microdischarge operating in helium at 100 Torr ($D_{cav}$ = 260 µm) for different dimensions of the limiting layer. The opening diameter of the limiting layer is 1 mm and its thickness is 60 µm or 1 mm.*

As already mentioned in the discussion of Figure 6b, the gas can reach temperatures as high as 600 K in the bulk of the micro-discharge for $I_d$ = 4 mA, with $D_{LIM}$ = 1 mm and $t_{LIM}$ = 60 µm. Most of the gas heating, initially transferred to the cathode layer by ion impact and by thermal contact with the hot gas (through a thermal boundary layer) is then transmitted to the dielectric layers (alumina and Kapton). By increasing the thickness of the limiting layer (Kapton) one can expect a more efficient heat transfer out of the gas. (All surfaces are held at a constant temperature of 300 K in the simulation and thus serve to lower the gas temperatures in the micro-discharge.) The calculated gas temperature distributions are shown in Figure 15 for three cases: (a) no limiting layer, (b) $t_{LIM}$ = 60 µm and (c) $t_{LIM}$ = 1 mm. The thermal contact between the micro-discharge and the dielectric as well as metallic electrodes is over a larger surface area in case (c), thus promoting more efficient heat losses. The case (b) geometry is the worst case because the spreading of the micro-discharge on the cathode surface is limited and so is the contact to the dielectric cooling surface. As a result, the current density is higher than in case (a) and hence the gas temperature is higher too (up to 650 K).

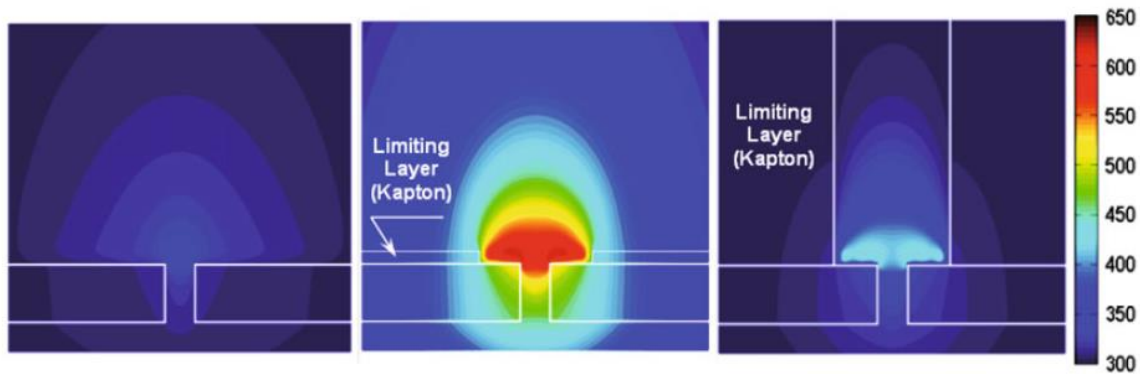

*Fig. 15. A zoom of the gas temperature (K) profiles in the cathode region simulated for a micro-discharge operating in helium at 100 Torr for $D_{LIM}$ = 1 mm. (a) $t_{LIM}$ = 0; (b) $t_{LIM}$ = 60 µm; (c) $t_{LIM}$ = 1 mm.*

# 4. Conclusion

Operating arrays microdischarges is problematic if the individual discharges function in the normal glow regime where the V–I characteristic is flat or even slightly negative. We have previously demonstrated [4] that it is possible to operate MHCDs in an abnormal regime with a positive V – I characteristic if the cathode surface area is limited by a thin dielectric layer on the cathode surface. Note that the Cathode Boundary Layer discharges (CBLs) discussed by Takano and Schoenbach [23] are another microdischarge configuration which functions in an abnormal mode by limiting the surface area available for the discharge to spread. A modelling study of the properties of CBLs was presented by Munoz-Serrano et al. [24] and will be published separately. Experimental measurements and simulation results reported here provide insights on the physical properties underlying the operation of micro-discharges with the cathode surfaces limited by a dielectric layer. First, we have demonstrated that in our hollow cathode geometry, no hollow cathode effect is detected, not unexpectedly as the cathode is quite thin (only 8 µm). Therefore, the breakdown voltage is a function of the interelectrode distance and is not influenced by any pendulum electrons. Second, we have demonstrated using both experiment and simulation that the micro-discharge can be operated in either a normal or an abnormal regime, depending on surface area available on the cathode. Limiting the cathode surface area has a large influence on the discharge properties – operating voltage, maximum current and V – I hysteresis. Reducing $D_{LIM}$ also has a strong influence on the cathode sheath thickness and on the gas temperature which was found to reach up to 600 K at 4 mA.